\documentstyle[12pt,epsf]{article}
\begin{document}
\def\lsim{\mathrel{\lower4pt\hbox{$\sim$}}\hskip-12pt\raise1.6pt\hbox{$<$}\;
}
\def\BAR{\bar}
\def\xba{\bar}
\def\fm{{\cal M}}
\def\fl{{\cal L}}
\def\ufs{\Upsilon(5S)}
\def\gsim{\mathrel{\lower4pt\hbox{$\sim$}}
\hskip-10pt\raise1.6pt\hbox{$>$}\;}
\def\ufour{\Upsilon(4S)}
\def\xcp{X_{CP}}
\def\ynotcp{Y}
\vspace*{-.5in}
\def\etap{\eta^\prime}

\def\uglu{\hskip 0pt plus 1fil
minus 1fil} \def\uglux{\hskip 0pt plus .75fil minus .75fil}

\def\slashed#1{\setbox200=\hbox{$ #1 $}
    \hbox{\box200 \hskip -\wd200 \hbox to \wd200 {\uglu $/$ \uglux}}}
\def\slpar{\slashed\partial}
\def\sla{\slashed a}
\def\slb{\slashed b}
\def\slc{\slashed c}
\def\sld{\slashed d}
\def\sle{\slashed e}
\def\slf{\slashed f}
\def\slg{\slashed g}
\def\slh{\slashed h}
\def\sli{\slashed i}
\def\slj{\slashed j}
\def\slk{\slashed k}
\def\sll{\slashed l}
\def\slm{\slashed m}
\def\sln{\slashed n}
\def\slo{\slashed o}
\def\slp{\slashed p}
\def\slq{\slashed q}
\def\slr{\slashed r}
\def\sls{\slashed s}
\def\slt{\slashed t}
\def\slu{\slashed u}
\def\slv{\slashed v}
\def\slw{\slashed w}
\def\slx{\slashed x}
\def\sly{\slashed y}
\def\slz{\slashed z}

\rightline{AMES-HET 01-13}
\rightline{BNL-HET-01/41}

\begin{center}

{\large\bf
Using Imprecise Tags of CP Eigenstates in $B_s$ and the Determination of
the CKM Phase $\gamma$
}

\vspace{.2in}

David Atwood$^{1}$\\
\noindent Department of Physics and Astronomy, Iowa State University, Ames,
IA\ \ \hspace*{6pt}50011\\
\medskip

Amarjit Soni$^{2}$\\
\noindent Theory Group, Brookhaven National Laboratory, Upton, NY\ \
11973\\
\footnotetext[1]{email: atwood@iastate.edu}
\footnotetext[2]{email: soni@bnl.gov}
\end{center}
\vspace{.15in}

\begin{quote}
{\bf Abstract:}
We consider the possibility of studying the CP properties of various $B_s$
decays at an electron-positron machine tuned to the $\ufs$.  Since decay
modes of the $B_s$ with definite CP are relatively rare, we suggest that
the use of more common modes which are not pure CP eigenstates may allow
the determination of the CKM phase $\gamma$. By studying the degree of
correlation between different decay modes at a $\ufs$ it is possible to
determine the degree of affinity of each decay mode to a CP eigenstate.
Once this is known, the correlation between a decay mode with a greater
affinity to a particular CP eigenstate with a mode such as $D_s^+K^-$
gives a determination of the phase $\gamma$.
\end{quote}

\newpage

$B$ factories currently running on the $\ufour$ resonance will shed light
on a number of important aspects of the Standard Model (SM), most
prominently CP violation through $B^0\to J/\psi K_s$ as well as a number
of other rare modes~\cite{bfac_general}.  To a large extent various rare
decay modes should confirm or refute the role of the SM in $B$ decays.
To gather a more complete understanding, however, it is desirable to also
study $B_s$ decays.

The remarkable success of the current B-factory programs opens up the
possibility that future colliders could be run at luminosities an order
of magnitude (or more) 
beyond those being achieved in the current program.  If
luminosities on the order of $10^{35}~cm^{-2}s^{-1}$ prove to be
achievable, in addition to the contributions to $B$ physics which will
result from the study of $\Upsilon(4s)\to B\xba B$, it may be worthwhile
to running such machines at the $\ufs$ peak to study $\ufs\to B_s B_s$
(and also $B_sB_s^*$ and $B_s^*B_s^*$)~\cite{bar_note}.

The yield of $B_s$ mesons in such a scenario can be roughly estimated from
existing data~\cite{csub}. The cross section for $\ufs$ production is
about $320~pb$~\cite{csub} while the branching ratio for $B_sB_s$ +
$B_s^*B_s$ + $B_s^*B_s^*$ was found to be about $0.3$; hence at a
luminosity of $10^{35}~cm^{-2}s^{-1}$ about $10^8$ such pairs would be
produced per year.  Although hadronic $B$ machines could produce similar
numbers of $B_s$ mesons, at an $e^+e^-$ collider the experimental
environment would be cleaner and, more significantly, the pairs produced
would be in a correlated CP eigenstate. In this paper, we consider how
this fact may be exploited to study the general properties of the decay of
$B_s$ eigenstates and probe the angle $\gamma$ of the CKM matrix.

A unique feature of the $B_s$ system is that the two eigenstates are
likely to have appreciably different inclusive and exclusive decay
rates~\cite{bs_life_diff}.  In particular, if there is a significant
difference between the total decay rates of the two eigenstates, it
logically follows that there should be even greater differences in some
exclusive decay channels.

As discussed in~\cite{xing} the fact that indirect CP violation in $B_s$
is small in the SM means that a $B_sB_s$ correlated CP state is an ideal
system to look for either large CP violating mixing from new physics or
direct CP violation in $B_s$ decay as in~\cite{falkpetrov} which we focus
on here.

In this paper, let us assume that indirect CP violation in the $B_s$ is
small so that the mass eigenstates are nearly CP eigenstates (we denote
these eigenstates $B_s^{(+)}$ and $B_s^{(-)}$). Thus, some decay modes,
such as $D_s^+D_s^-$ which are explicit CP eigenstates would only be
produced in the decay of the corresponding $B_s$ state. Since these modes
have small branching ratios, if the width difference, $\Delta\Gamma$, is
large the bulk of $\Delta\Gamma$ must be accounted for by CP
non-eigenstates.

The $\ufs$ is an ideal arena to study such effects
because:

\begin{itemize}
\item
When $\ufs\to B_s B_s$ the two mesons are in opposite CP
eigenstates; similar correlations are true in the final states through the
channels $B_sB_s^*$ and $B_s^*B_s^*$.
\item
In the SM, the CP eigenstates closely approximate the mass
eigenstates (unlike the flavor eigenstates), thus the correlation persists
until the mesons decay.
\end{itemize}

It has been pointed out by Falk and Petrov~\cite{falkpetrov} that this
correlation may be used to determine the angle $\gamma$ of the CKM
matrix.  In the method of Falk and Petrov (FP), one observes decays of
the form $\ufs\to B_s B_s$ and determines the probability that the
decay of one of the $B_s$ to a CP eigenstate is correlated to the decay
of the other $B_s$ to states such as $D_s^\pm K^\mp$ with a $c\bar u s
\bar s$ (or charge conjugate) quark content.  From such measurements, one
obtains the branching fraction of each of the CP eigenstates to the final
states $D_s^\pm K^\mp$. Clearly then, a difference between the branching
ratios of $B_s^{(+)}\to D_s^+ K^-$ and $B_s^{(+)}\to D_s^- K^+$ would
violate CP (and likewise for $B_s^{(-)}$).  In the SM this CP violation
is proportional to the CKM phase $\gamma$ and in fact this angle may be
extracted from such data.

In this letter we generalize the FP method to include the use of both
inclusive or exclusive CP non-eigenstate decays. In this way, one hopes to
overcome the problem that the branching ratio to explicit CP eigenstates
is small while giving up perfect knowledge of the CP state in $B_s\to
D_s^\pm K^\mp$.  In particular, this strategy is guaranteed to be more
effective than the FP method if $\rho\equiv\Delta\Gamma/\Gamma$ is large
(e.g. $\sim 15\%$). This follows from the fact that $\rho$ must be
entirely accounted for by the inclusive $c\bar c s\bar s$ channel. The
relatively large branching ratio of this channel together with the large
decay rate difference can lead to a determination of $\gamma$ with 
about a factor of three fewer number of pairs of $B_s$ 
than is needed with the FP method.  Even if the width
difference is small ($\sim 5\%$), the two approaches may be comparable,
and, in addition, by weighting the individual decay modes, the sensitivity
will be improved although the amount of this improvement cannot easily be
determined a priori.

A key ingredient in this analysis is the affinity that a given inclusive
or exclusive decay mode has to the CP eigenstates. The correlations in
$\ufs\to B_sB_s$ allows us to determine this as well. In fact, assuming
that $\Delta\Gamma$ is largely due to $c\bar c s \bar s$ we obtain as a
byproduct of such studies a determination of $\Delta\Gamma/\Gamma$ with
time independent data.

In this paper, we will discuss our generalization of the FP method in the
context of a $\ufs$ collider with luminosity of $\sim 10^{35}
cm^{-2}s^{-1}$ as a means to determine $\gamma$ using general methods of
CP tagging.  Key to this is the use of $\ufs$ to calibrate CP affinity for
various $B_s$ decay modes. As another important application of these
methods we will suggest how to determine $\Delta\Gamma/\Gamma$ using time
independent correlations in $\ufs$ decay.

Let us first turn our attention to the correlations that exist in $B_s$
production at $\ufs$. There are three decay channels of the $\ufs$ which
lead to a $B_s$ pair: $\ufs\to B_sB_s$, $B_sB_s^*$ and $B_s^*B_s^*$.  
Since the $B_s^*$ promptly decays through $B_s^*\to B_s\gamma$, in all
these cases a $B_s$ pair is produced with $0-2$ associated photons.  The
CP correlation of the $B_s$ pair is different for the different production
channels so it is crucial for any experiment considered here to determine
the number of photons present.

Let us define $\chi$ to be the product of the CP eigenvalues of the two
$B_s$ mesons.  If the decay channel is $\ufs\to B_s B_s$ or $\ufs\to B_s^*
B_s^*$ (ie 0 or 2 photons), then it is clear that CP states of the two
mesons will be opposite so $\chi=-1$.

In the case of $B_sB_s^*$, the pair need not be a purely $\chi=\pm 1$
state.  The kinematics of the decay, however, dictate it is largely in the
$\chi=+1$ (correlated)  state. To see this, let us write the form of the
$\ufs \bar B_s B_s^*$ coupling determined by parity and Lorentz
invariants:

\begin{eqnarray}
\propto~
\epsilon_{\mu\nu\rho\sigma}E_\Upsilon^\mu p_\Upsilon^\nu E_1^\rho p_1^\sigma
\end{eqnarray}

\noindent where $E_\Upsilon$ is the polarization of the $\ufs$,
$p_\Upsilon$ is the 4-momentum of the $\ufs$ and $E_1$ and $p_1$ are the
polarization and 4-momentum of the $B_s^*$. Likewise the coupling of the
$B_s^*$ to the $\gamma B_s$ is:

\begin{eqnarray}
\propto~
\epsilon_{\mu\nu\rho\sigma}E_\gamma^\mu p_\gamma^\nu E_1^\rho p_1^\sigma
\end{eqnarray}

If we define $R$ to be the proportion of $\chi=-1$, $B_sB_s$ pairs coming
from the $B_sB_s^*$ channel:

\begin{eqnarray}
R={
Br(\ufs\to  B_sB_s\{\chi=-1\} +\gamma)
\over
Br(\ufs\to B_s B_s^*)
}.
\end{eqnarray}

This ratio can be shown to be small by expanding in the small mass
difference between $B_s$ and $B_s^*$ and the difference between $m_{\ufs}$
and $m_{B_s}+m_{B_s}^*$:

\begin{eqnarray}
R={3\over 8} ru + {1\over 64}r^2(106 v^2-36 uv-9u^2) +O(r^3u^3,r^3v^3)
\end{eqnarray}

\noindent where $v=(m_{\ufs}-m_{B_s}-m_{B_s}^*)/m_{B_s}$;
$u=(m_{B_s}^*-m_{B_s})/m_{B_s}$ and $r=u/v$. Numerically, this gives
$R\approx 0.002$ and so for all practical purposes such a $B_s$ pair is in
the $\chi=+1$ state. In summary then, $\chi=+1$ for $\ufs\to B_s
B_s+\gamma$ and $\chi=-1$ for $\ufs\to B_s B_s$ or $B_s B_s+2\gamma$.

Let us now consider an exclusive or inclusive final state $A$.
We define the CP affinity $\alpha_A$ to be

\begin{eqnarray}
\alpha_A=
{Br(B_s^{(+)}\to A)-Br(B_s^{(-)}\to A)
\over
Br(B_s^{(+)}\to A)+Br(B_s^{(-)}\to A)}
\end{eqnarray}

\noindent 
So that $\alpha_A=\pm 1$ for CP eigenstates with eigenvalue $\pm 1$ but
takes on some intermediate value for all other states.

Assuming that CP is conserved in $B_s$ mixing, the correlated branching
ratio for a $B_s$ pair to decay to a combination of final states $A$ and
$X$ is thus:

\begin{eqnarray}
Br(B_sB_s\to AX)=
2 {\overline Br}(A){\overline Br}(X)(1+\chi \alpha_A\alpha_X)
\label{correl}
\end{eqnarray}

\noindent where ${\overline Br} (X)=(Br(B_s^{(+)}\to X) + Br(B_s^{(-)}\to
X))/2$. Therefore the observation of $Br(B_sB_s\to AX)$ tells us
$\alpha_X$ if ${\overline Br}(A)$, ${\overline Br}(X)$ and $\alpha_A$ are
known. In our generalization of the FP method to determine $\gamma$ we
will assume that for some final state, $A$, $\alpha_A$ is known.  Once
$\alpha_A$ is known, then we can use eqn.~(\ref{correl}) to determine
$\alpha_X$ for the cases $X=D^\pm K^\mp$.

Consider now the amplitude for various $B_s$ states to decay to $D^\pm
K^\mp$.  The amplitudes for flavor eigenstates may be written as:

\begin{eqnarray}
     A_1&=&A(     B_s\to D_s^-K^+)=a_1 e^{i\delta_1}
~~~~(\xba b \to \xba c)           \\
     A_2&=&A(\xba B_s\to D_s^-K^+)=a_2 e^{i(\delta_2-\gamma)}
~~~~(b\to u)   \\
\xba A_1&=&A(     B_s\to D_s^+K^-)=a_2 e^{i(\delta_2+\gamma)}
~~~~(\xba b \to \xba u)    \\
\xba A_2&=&A(\xba B_s\to D_s^+K^-)=a_1 e^{i\delta_1}
~~~~(b\to c)             
\end{eqnarray}

\noindent where $\delta_{1,2}$ are strong phases
and the quark level processes are indicated for 
each amplitude in parenthesis.  
We can now write
the
amplitudes for the decay of CP eigenstates:

\begin{eqnarray}
A_\pm 
= 
A(B_s^{(\pm)}\to D_s^-K^+) 
=
{A_1\pm A_2\over\sqrt{2}}
\nonumber\\
\xba A_\pm 
=
A(B_s^{(\pm)}\to D_s^+K^-) 
= 
{\xba A_1\pm \xba A_2\over\sqrt{2}}
\end{eqnarray}

For now, let us assume that we know the magnitude of the decays from
flavor eigenstates, $|A_1|$ and $|A_2|$. If we can determine $|A_\sigma|$
for either $\sigma=+$ or $\sigma=-$, then the FP method extracts $\gamma$
by defining:

\begin{eqnarray}
a={2|A_\sigma|^2-|A_1|^2-|A_2|^2\over 2~|A_1|~|A_2|}
=\sigma\cos(\delta-\gamma)
\nonumber\\
\xba a
={2|\xba A_\sigma|^2-
|\xba A_1|^2
-|\xba A_2|^2
\over 
2~|\xba A_1|~|\xba A_2|}
=\sigma\cos(\delta+\gamma)
\label{adef}
\end{eqnarray}

\noindent
where $\delta=\delta_2-\delta_1$.
The angle $\gamma$
may thus be obtained from 
the relation:

\begin{eqnarray}
\gamma 
= 
(\pm \cos^{-1} a
\pm \cos^{-1} \xba a)/2
~~~~~~~~
({\rm mod}~\pi)
\label{gammaeqn}
\end{eqnarray}

\noindent
up to an 8 fold ambiguity 
in $\gamma$
(each of the $\pm$ are independent and there is a 2 fold
${\rm mod}~\pi$  ambiguity).

The ingredients we need to carry out this program are $|A_\sigma|$, $|\xba
A_\sigma|$, $|A_1|$ and $|A_2|$. The key contribution which can be made at
a $\ufs$ collider is the determination of $|A_\sigma|$. This can be done
through the use of eqn.~(\ref{correl})  since

\begin{eqnarray}
|A_\sigma|^2
\propto \Gamma(B^{(\sigma)}\to D_s^- K^+)
= (1+{\sigma\over 2}\rho)(1+\sigma \alpha_{D_s^-K^+})\xba\Gamma
\nonumber\\
|\xba A_\sigma|^2
\propto \Gamma(B^{(\sigma)}\to D_s^+ K^-)
= (1+{\sigma\over 2}\rho)(1+\sigma \alpha_{D_s^+K^-})\xba\Gamma
\end{eqnarray}

\noindent where $\xba\Gamma = \left (\Gamma(B^{(+)}+\Gamma(B^{(-)})\right
)/2$.

Although it is in principle possible to determine $|A_1|$ and $|A_2|$ by
time dependent studies at a hadronic collider, 
these
quantities 
cannot be determined at a $\ufs$ collider since the
flavor of the $B_s$ at the time of decay would need to be tagged.  One can,
however, measure:

\begin{eqnarray}
R_0={1\over 2} (|A_1|^2+|A_2|^2)
\end{eqnarray}

\noindent
which is proportional to the rate of untagged $B_s$ decay to $D_s^-K^+$.
It is the ratio:

\begin{eqnarray}
\nu={|A_2|\over |A_1|}
\end{eqnarray}

\noindent
that is problematic.

One can, however, estimate this ratio (of quark level $b\to u$ to $b\to
c$ transitions) assuming spectator dominance by considering
analogous branching ratios where the spectator quark is replaced by a $u$
or $d$. Thus:

\begin{eqnarray}
\nu
=
{A(B_s\to D_s^+K^-)\over A(B_s\to D_s^-K^+)}
\approx
{A(B_d\to D_s^+\pi^-)\over A(B_d\to D^-K^+)}
\approx \sqrt{2}
{A(B_u\to D_s^+\pi^0)\over A(B_u\to D^0K^+)}
\end{eqnarray}

\noindent 
where much of the uncertainty cancels in the ratio.

Thus, rewriting eqn.~(\ref{adef}) in terms of $R_0$ and $\nu$ it 
becomes:

\begin{eqnarray}
a= \left ( { 1+\nu^2\over 2\nu}\right )
   \left ({  |A_\sigma|^2 -R_0 \over R_0 }\right)
=\sigma\cos(\delta-\gamma)
\nonumber\\
\xba a= \left ( { 1+\nu^2\over 2\nu}\right )
        \left ({  |\xba A_\sigma|^2 -R_0 \over R_0 }\right)
=\sigma\cos(\delta+\gamma)
\end{eqnarray}

\noindent
where the dependence on the problematic  factor $\nu$ has been 
factored out in front.

As stated before, in order to carry out the above procedure, it is
necessary that some mode $A$ with a known value of $\alpha_A$ be
available. Let us therefore consider the determination of $\alpha_A$.  
Again, using the CP correlations of the $\ufs$ we will consider two
experimental procedures to extract this quantity:

\begin{enumerate}
\item
Determine the correlation in $\ufs\to A A$.
\item
Determine the correlation in $\ufs\to A X$. Here $X$ is a state for which  
$\alpha_X$ is already known.
\end{enumerate}

Assuming that $B_sB_s$, $B_s^*B_s^*$ can be distinguished from $B_s^*B_s$,
let us now consider the relative merits of these two methods. As we shall
see, the preferred method will depend on the value of $\alpha_A$ and
$\alpha_X$ as well as the corresponding branching ratios (which we denote
$B_A$ and $B_X$).
We also denote by $b$ the portion of events that have $\chi=-1$, i.e. 
$b=(\# B_sB_s + \# B_s^*B_s^*)/({\rm all~B_sB_s~states})$.

Consider first the use of self-correlation to determine $\alpha_A$. The
total statistical error, in the determination of $\alpha_A$ one obtains by
combining data from all three $B_sB_s$ modes is given by:

\begin{eqnarray}
(\Delta \alpha_A(AA))^2
=
(4\alpha_A^2B_A^2N)^{-1} \left[{1-\alpha_A^4\over 1-(1-2b)\alpha_A^2}\right]
\end{eqnarray}

On the other hand, if we use the mode $X$ where $\alpha_X$ is assumed to
already be known, then the error in determination of $\alpha_A$ is:

\begin{eqnarray}
(\Delta \alpha_A(AX))^2
=
(2B_AB_X\alpha_X^2 N)^{-1}
\left[{
1-\alpha_A^2\alpha_X^2
\over
1-\alpha_A\alpha_X(1-2b)
}\right]
\end{eqnarray}

Thus using the second method will be statistically more favorable if 
\begin{eqnarray}
{
B_A\alpha_A^2
\over 
B_X\alpha_X^2
}
>
2\left[{
(1-\alpha_A^4)(1-\alpha_A\alpha_X(1-2b))
\over
(1-\alpha_A^2\alpha_X^2)(1-\alpha_A^2(1-2b))
}\right]
\end{eqnarray}

There are two classes of possible CP tagging states which should
initially be considered.  First of all there are those which are
themselves CP eigenstates and thus are pure CP tags.  In such cases
$\alpha=\pm1$ is given in the limit that there is no CP violation in the
mixing of $B_s$ and no direct CP violation in the decay to that mode.  
Some prominent decay modes of this type are $D_sD_s$ and $J/\psi
\eta^{(\prime)}$. Extrapolating from known $B$ decays, the branching
ratios to these states should roughly be $Br(B_s\to D_s^+D_s^-)\approx
Br(B^0 \to D^- D_s^+)\approx 0.8\%$ while $Br(B_s \to J/\psi
\eta^{(\prime)})\approx Br(B^0 \to J/\psi \eta^{(\prime)})\approx
0.15\%$.

To make up for the relatively small branching ratios of 
the pure exclusive CP eigenstates considered above, it is useful to
consider impure inclusive and exclusive states ($|\alpha|\leq 1$), in
particular
states with quark content $c\bar c s \bar s$.
Some states of this type are $D_s D_s^*$, $J/\psi~\phi$,
$J/\psi+X$. For any one of these states, it is hard to predict what
$\alpha$ will be, and so it would have to be determined experimentally.

Indeed the set of states consisting inclusively of all $c\xba c s \xba s$
(which we will denote $X_{ccss}$) is probably the most promising.  A
number of separate calculations\cite{bs_life_diff} indicate that the
width difference ($\rho$) could be as large as $\rho\sim 15\%$ and since
this width difference is almost entirely accounted for by $c\xba cs\xba
s$ states,

\begin{eqnarray}
\alpha_{ccss}={\rho\over 2}   {1- B_{ccss}\over B_{ccss}}.
\label{rhoeqn}
\end{eqnarray}

\noindent
Thus, if $\rho=0.15$ and assuming that the branching ratio $Br(B_s\to
X_{ccss})\approx 0.24$ as in the similar $B$ decay, we obtain
$\alpha\approx 0.24$.

The analyzing power defined by ${\cal A}(A)=\alpha_A^2 B_A$ quantifies the
usefulness of a given mode in CP asymmetry studies. Thus for $\rho=.15$,
${\cal A}(X_{ccss})=2.5\%$. Clearly this is superior to all of the CP
eigenstate modes (where ${\cal A}=B_A$) by a factor of at least 3.

To experimentally determine $\alpha(X_{ccss})$, the self correlation
method would likely be the best. If $N=10^8$ (the number of $B_s$ pairs)
then the statistical error in $\alpha$ would be $0.07\%$ using the
formalism above.  In practice using this method it is the subset of
$X_{ccss}$ which is accepted by the detector which will have $\alpha$
directly determined. In addition, the data may allow the identification of
a subset of $X_{ccss}$ which might have a larger value of ${\cal A}$.

%
%
%
%

Indeed, this provides a time independent method of determining the $B_s$
width difference.  If we assume that the entire width difference is due to
final states of the form $c\bar c s \bar s$ then $\rho$ can be determined
through the relation eqn.~(\ref{rhoeqn}) if $\alpha_{ccss}$ and $B_{ccss}$
have been experimentally determined.

Taking the number of $B_s$ pairs, $N=10^8$, let us now estimate the
statistical error one might expect in $\sin 2\gamma$. First, we need an
estimate of $R_0$. These branching ratios are not as yet measured but by
comparison with the corresponding $B^+$ decay:

\begin{eqnarray}
R_0\approx {1\over 2}\sin^2\theta_c Br(B^+\to D^0\pi^+)\approx 1.3\times
10^{-4}
\end{eqnarray}

\noindent and 

\begin{eqnarray}
\nu\approx 
\left | {V_{cb}V_{us}\over V_{cs}V_{ub}}\right |
\approx
2.6
\end{eqnarray}

For the purposes of a numerical estimate, we will assume that  
$\gamma=90^\circ$ and $\delta=0^\circ$. Taking ${\cal A}=2.5\%$
the statistical error in $\sin 2\gamma$ is thus:

\begin{eqnarray}
\Delta (\sin2\gamma) = 0.09
\end{eqnarray}

\noindent
In general, we could reduce the error further by combining data from
similar decay modes
such as
$B_s^*K$ and 
$B_sK^*$.

Another use for having an enhanced ability to tag CP eigenstates is to
test for CP violation in the mixing of the $B_s$ at a hadronic B machine.  
If, as in the kaon system, we denote the CP eigenstates as:

\begin{eqnarray}
|B_1>={(1+\epsilon)|B_s> + (1-\epsilon)|\xba B_s>\over
\sqrt{2(1+\epsilon^2)} }
\nonumber\\
|B_2>={(1+\epsilon)|B_s> - (1-\epsilon)|\xba B_s>\over
\sqrt{2(1+\epsilon^2)} }
\end{eqnarray}

\noindent Assuming the meson can be tagged at the point of creation as
either a $B_s$ or $\xba B_s$ then, following the formalism
of~\cite{quinn_pdb} the time dependent rate of decay to a $CP=+$ or a
$CP=-$ state to leading order in $\epsilon$ is:

\begin{eqnarray}
\Gamma(B_s(t)\to X_+)
&=&
{1\over 2}
e^{-\xba \Gamma t/2}
(C_\Gamma - 2(1-2\epsilon_r)S_\Gamma + 4\epsilon_r C_M + 4\epsilon_iS_M)
\nonumber\\
\Gamma(\xba B_s(t)\to X_+)
&=&
{1\over 2}
e^{-\xba \Gamma t/2}
(C_\Gamma + 2(1-2\epsilon_r)S_\Gamma - 4\epsilon_r C_M + 4\epsilon_iS_M)
\nonumber\\
\Gamma(B_s(t)\to X_-)
&=&
{1\over 2}
e^{-\xba \Gamma t/2}
(C_\Gamma + 2(1-2\epsilon_r)S_\Gamma + 4\epsilon_r C_M - 4\epsilon_iS_M)
\nonumber\\
\Gamma(\xba B_s(t)\to X_-)
&=&
{1\over 2}
e^{-\xba \Gamma t/2}
(C_\Gamma - 2(1-2\epsilon_r)S_\Gamma - 4\epsilon_r C_M- 4\epsilon_iS_M)
\nonumber\\
\end{eqnarray}

\noindent
where $\xba \Gamma = (\Gamma_1 + \Gamma_2)/2$, 
$C_\Gamma=\cosh(\Delta\Gamma t/2)$,
$S_\Gamma=\sinh(\Delta\Gamma t/2)$,
$C_M=\cos(\Delta M t)$,
$S_M=\sin(\Delta M t)$
and
$\epsilon=\epsilon_r+i\epsilon_i$.

As can be seen, the oscillating terms are proportional to $\epsilon_r$ and
$\epsilon_i$. 
In the Standard Model, it is estimated that
$\epsilon=O(10^{-3})$ so 
the observation of
this form of CP violation would be indicative of
physics beyond the SM.

Of course in the discussion above we have assumed that CP is preserved in
the mixing in order to correlate the states of the $B_s$ mesons from the
$\ufs$. This assumption is reasonable since the time integrated
probability of a $B_s^+$ state decaying to a $CP=-$ state is $\propto
\epsilon^2$. 
In particular, neglecting terms $\propto \Delta \Gamma/\Delta
M$,

\begin{eqnarray}
{
\Gamma(B_s^{(+)}\to CP-)
\over
\Gamma(B_s^{(+)}\to CP+)
}
=
(1+\Gamma_1/\Gamma_2)\left | \epsilon \right | ^2
\end{eqnarray}

\noindent
and thus negligible for small $\epsilon$.

In conclusion, a collider run at the $\ufs$ resonance has a number of
potential applications due to the correlation of the CP states of the
$B_s$ meson pairs it produces. In general, we can explore the differences
in the decay rates to inclusive and exclusive final states either through
self correlation or correlation with a known CP eigenstate.  In
particular, if $\alpha_{ccss}$ is measured, then we can determine
$\rho\equiv\Delta\Gamma/\Gamma$ through time independent data. In the
case of $D_s^\pm K^\mp$ we can use this data to determine the CKM phase
$\gamma$ using the generalization of the FP method.  Note that for FP and
its generalization we still need the parameter $\nu$, however, this may be
estimated from theoretical arguments or be determined at a hadronic
machine. We can also use the differences between the decays of the CP
eigenstates to look for CP violating mixing effects at a hadronic
machine.

\bigskip

We 
thank
Adam Falk for useful discussions 
concerning the ambiguities in eqn.~(\ref{gammaeqn}).
This research was supported in part by US DOE Contract Nos.\
DE-FG02-94ER40817 (ISU); DE-AC02-98CH10886 (BNL).


\newpage





\begin{thebibliography}{99}


\bibitem{bfac_general}
J.~T.~Seeman {\it et. al.}, Fourth International Workshop on B Physics and
CP Violation, Isi-Shema, Japan (Feb. 19 2001);
S.~Kurokawa, Fourth International Workshop on B Physics and
CP Violation, Isi-Shema, Japan (Feb. 19 2001).


\bibitem{bar_note}
In this paper we will often consider $B_s$ mesons in their CP
eigenstates. We will therefore generally use the notation $B_s$ in the
generic sense
to mean any linear combination of 
$|B_s\rangle$ and 
$|\bar B_s\rangle$. 




\bibitem{csub}
C.~Yanagisawa {\it et al}, Phys.\ Rev.\ Lett.\ {\it 66}, 2436, (1991);
J.~Lee-Franzini {\it et al}, Phys.\ Rev.\ Lett.\ {\it 65}, 2947, (1990).  





\bibitem{bs_life_diff}
A.~Datta, E.~A.~Paschos and U.~Turke,
Phys.\ Lett.\ B {\bf 196}, 382 (1987);
M.~Beneke, G.~Buchalla and I.~Dunietz,
Phys.\ Rev.\  {\bf D54}, 4419 (1996);  
M.~Beneke, G.~Buchalla, C.~Greub, A.~Lenz and U.~Nierste,
Phys.\ Lett.\  {\bf B459}, 631 (1999);
S.~Hashimoto, K.~I.~Ishikawa, T.~Onogi and N.~Yamada,
Phys.\ Rev.\ D {\bf 62}, 034504 (2000);
D.~Becirevic {\it et. al.}
Eur.\ Phys.\ J.\ C {\bf 18}, 157 (2000);
M.~Beneke and A.~Lenz,
J.\ Phys.\ G {\bf 27}, 1219 (2001).



\bibitem{xing}
Z.-z.~Xing,
Eur.\ Phys.\ J.\ C {\bf 4}, 283 (1998).





\bibitem{falkpetrov}
A.~F.~Falk and A.~A.~Petrov,
Phys.\ Rev.\ Lett.\ {\bf 85}, 252 (2000);
A.~F.~Falk,
Phys.\ Rev.\ D {\bf 64} (2001) 093011;
Z.-z.~Xing,
Phys.\ Lett.\ B {\bf 488}, 162 (2000).









\bibitem{quinn_pdb}
H.~Quinn and A.~I.~Sanda,
article in {\it Review of Particle Properties}, 
D.~E.~Groom {\it et al}, Eur. Phys. J. C. , {\bf C15}, 1 (2000).




\end{thebibliography}
\end{document}